\documentclass[aps,pra,showpacs,preprint]{revtex4}
\usepackage{amssymb}
\usepackage{graphicx}
\usepackage{amsmath}
\usepackage{subfigure}
\usepackage{epsfig}
\usepackage{multirow}
\usepackage{float}
\usepackage{color}
\usepackage{lscape}
\usepackage{pdflscape}
\usepackage[figuresright]{rotating}

\begin{document}

\title{Finite-field calculation of the polarizabilities and hyperpolarizabilities of Al$^{+}$}

\author{Yan-mei Yu$^{1}$ \footnote[1]{E-mail: ymyu@aphy.iphy.ac.cn}, Bing-bing Suo$^{1,2}$\footnote[2]{E-mail: bingbing.suo@gmail.com}, and Heng Fan$^{1}$\footnote[3]{E-mail: hfan@aphy.iphy.ac.cn}}
\address{$^1$Beijing National Laboratory for Condensed Matter Physics, Institute of Physics, Chinese Academy of Sciences, Beijing 100190,China}
\address{$^2$Institute of Modern Physics, Northwest University, Xi'an, Shanxi 710069, China}
\date{\today}
\begin{abstract}
In this study, accurate static dipole polarizability and hyperpolarizability are calculated for Al$^+$ ground state $3s^{2}\ ^{1}S_{0}$ and excited state $3s3p\ ^{3}P_{J}$ with $J=0, 1, 2$. The finite-field computations use energies obtained with the relativistic configuration interaction approach and the relativistic coupled-cluster approach. Excellent agreement with previously recommended values is found for the dipole polarizability of Al$^{+}$ ground state $3s^{2}\ ^{1}S_{0}$ and excited state $3s3p\ ^{3}P_{0}$ as well as the hyperpolarizability of the ground state $3s^{2}$ $^{1}S_{0}$. The recommended values of the dipole polarizability of the Al$^{+}\ 3s3p\ ^{3}P_{1}$ and $^{3}P_{2}$ and the hyperpolarizability of Al$^{+}\ 3s3p\ ^{3}P_{0}$, $^{3}P_{1}$, and $^{3}P_{2}$ are also given. The impacts of the relativity and spin-orbit coupling are elucidated by analyzing the angular momentum dependence of the dipole polarizability and the hyperpolarizability and comparing the fully and scalar relativistic calculated data. It is shown that the impact of the relativity and spin-orbit coupling are small for the dipole polarizability but become significant for the hyperpolarizability. Finally, the black-body radiation shifts contributed by the dipole polarizability and hyperpolarizability respectively are evaluated for transitions of Al$^{+}\ 3s^{2}$ $^{1}S_{0}$ to $3s3p\ ^{3}P_{J}$ with $J=0, 1, 2$.
\end{abstract}


\maketitle

\section{Introduction}
The polarizabilities and hyperpolarizabilities are very useful quantities in many areas of atomic and molecular physics. The recent advance in development of the atomic optical clock has elevated great interest in the atomic polarizabilities and hyperpolarizabilities. The dipole polarizability determines the first-order response of the trapped atom or ion under the external perturbation, such as the electric field generated by the electrodes of an ion trap or the probe optical field, which brings the energy shifts that are main contributions to the frequency uncertainty budget of the atomic optical clock \cite{Itano-JRNIST-2000,Rosenband-arXiv-2000,Mitroy-JPB-2010,Safronova-IEEE-2012}. The higher-order response of atoms or ions to the applied electric field also contributes to the energy shift of the optical frequency standards, being small but not necessarily negligible \cite{Arora-PRA-2012, Porsev-PRA-2006}. Increasing order gives more accurate estimates of the polarization energy shift and the associate uncertainty. There already are a plenty of articles about the dipole polarizabilites of atoms and ions, most of which are about the ground state or monovalent system \cite{Lim-PRA-2004, Fleig-PRA-2005, Lupinetti-JCP-2005, Miller-CRC-2007, Thiefelder-PRA-2008, Schwerdtfeger-web-2012, Sahoo-PRA-2012,Porsev-PRA-2012}. The angular moment resolved dipole polarizability of the divalence systems and the hyperpolarizability remain very scarce for the majority of atoms or ions.

One important application of the highly accurate data of polarizabilities is to estimate the energy shifts in optical clock. As one of the highly accurate atomic clock to date \cite{Chou-PRL-2010}, the Al$^{+}$ optical clock, based on $3s^{2}\ ^1S_0$ $\rightarrow$ $3s3p\ ^3P_0$ transition, has attracted great interest in study of the polarizability properties of Al$^{+}$  \cite{Mitroy-EJPD-2009,Kallay-PRA-2011,Safronova-PRL-2011}. The coupling between the angular and spin momenta determines the multiple structure of the Al$^{+}\ 3s3p$ state, where in addition to $^3P_0$, there are the other two energetically higher lying states, $^3P_1$ and $^3P_2$. The polarizabilities of the $P$-state is dependent on the different $J$ components, for example, Fleig has studied the group-13 atoms, which has shown that the $J=1/2$ state differs from that of the $J=3/2$ as it directly depends on the spatial distribution of the electron density and also the mixing of spin and spatial degrees of freedom \cite{Fleig-PRA-2005}.

The relation between the polarizabilities corresponding to different $J$ components can be determined through basic vector algebra \cite{Mitroy-JPB-2010,Schwerdtfeger-web-2012,Angel-PRPSA-1968}. For the heavy elements, the spin splitting becomes very large and therefore the difference contributed by the spatial distribution of the electron density will become more pronounced. Such influence comes from the scalar relativity, the spin-orbit coupling and their combination, causing the possible deviation from the relationship derived by the basic vector algebra. Beside, the dipole polarizability and the hyperpolarizability are probably of different sensibility to the relativity. Therefore, it is important to resolve the polarizabilites for all $J$ components and their magnetic $M_J$ components directly, where $M_J$ is the projection onto the field axis, in order to understand the impacts of the relativity and the spin-orbit coupling on polarizabilities.

In the investigation of the dipole polarizability and hyperpolarizability, the finite-field (FF) method can provide reliable data if the field-dependent energies are calculated with high precision \cite{McLean-JCP-1967,Thakkar-JCP-1988}. Recently, the FF method has been implemented in the relativistic configuration interaction (CI) and the coupled cluster (CC) methods based on the four-component Dirac-Hatree-Fock (DHF) calculation \cite{Lim-PRA-2004, Fleig-PRA-2005, Thiefelder-PRA-2008, Kallay-PRA-2011}. The fully relativistic calculation allows us resolving electron states by total angular moment $J$, thus $J$-dependent properties can be obtained directly.

In the previous calculations, because the $^3P_0$ component is directly involved in the optional clock transition, most calculations are concentrated on this state, for example, Mitroy, et al. and Kallay, et al. have provided the dipole polarizability data of Al$^{+}\ 3s3p\ ^3P_0$, while the components with $J$=1 and 2 remain not available \cite{Mitroy-EJPD-2009, Kallay-PRA-2011}. Safronova, et al. have provided the dipole polarizability data of $nsnp\ ^3P_0$ of B$^+$, Al$^{+}$, In$^{+}$, Tl$^+$, and Sr \cite{Safronova-PRL-2011,Safronova-PRA-2012,Safronova-PRA-2013-Sr}. Cheng and Mitroy have done calculations on B$^+$ and Ga$^+$ \cite{Chen-PRA-2012, Chen-JPB-2013}. The polarizability data for the other $P_J$ states is scare for long time. Cheng, et al., have calculated the dipole polarizability of $nsnp$ $^3P_1$ state of Be, Mg, Ca, Sr atoms \cite{Cheng-PRA-2013}.

In the present work, we investigate Al$^{+}$ in order to give the $J$-resolved dipole polarizability and hyperpolarizability by using FF method. The field-dependent energies are obtained by using the relativistic CI calculation and the relativistic CC calculation. In addition to the dipole polarizability of $3s^{2}\ ^{1}S_{0}$ and $3s3p\ ^{3}P_{0}$ and the hyperpolarizability of $3s^{2}$ $^{1}S_{0}$, which is in excellent consistence with the previously recommended values \cite{Mitroy-EJPD-2009,Kallay-PRA-2011,Safronova-PRL-2011,Archibong-PRA-1991}, we also give the recommended values of the dipole polarizability of the $3s3p\ ^{3}P_{1}$ and $^{3}P_{2}$ and the hyperpolarizability of $3s3p\ ^{3}P_{J}$ with $J=0,1,2$. The difference in the dipole polarizability and hyperpolarizability for the different component of $3s3p\ ^{3}P$ state are studied, including the directional dependence by resolving the property in the azimuth projection $M_J$ substrates of the $J$ state, i.e., the anisotropy component. The impact of the relativistic effect on the dipole polarizability and the hyperpolarizability are elucidated by analogizing the $J$-dependence of such properties. The sole effect of the spin-orbit coupling on the polarizability and anisotropy components is determined by comparing the spin-dependent and the spin-free CI calculated data. Finally, the black-body radiation shifts by contributions from the dipole polarizability and hyperpolarizability are evaluated respectively for the transitions of Al$^{+}\ 3s^{2}\ ^{1}S_{0}$ to $3s3p\ ^{3}P_{J}$ with $J=0,1,2$.

\section{computational method}
The change in energy of an atom or ion upon introduction of a static, homogeneous, axially symmetric field $F_z$ is given by
\begin{equation}
\Delta E_{d}(F_z)=-\bar{\alpha} F_z^2/2-\bar{\gamma} F_{z}^4/24-\cdots,
\end{equation}
where $\bar{\alpha}$ is the dipole polarizability and $\bar{\gamma}$ is the hyperpolarizability. We apply the field in the $z$ direction, which allows us to retain a rotational axis, and therefore our symmetry choice is the double group $C^*_{2v}$ in Dirac-Hartree-Fock calculation and $C^*_2$ in the relativistic CI and the relativistic CC calculations. The $\bar{\alpha}$ and $\bar{\gamma}$ are obtained by fitting $\Delta E_q(F_z)$ versus $F_z$ with a 4th-order polynomial relationship. The field-dependent energies are calculated in an electric fields range $F_z$=[0, 0.0045] a.u with 0.0005 a.u. interval. Arbitrary more than four sample points are taken for fitting in order to check and remove the dependence of the studied properties on sampling. The reliable finite-field procedure depends on precise energies, where we converges the energies to $10^{-10} $ Hartree.

The field-dependent energy is calculated on the different level of theories, i.e., the spin-free CI calculation (implemented by LUCITA module in DIRAC package \cite{Dirac}), the spin-dependent CI calculation (implemented by KRCI module of DIRAC package), and the spin-dependent CC calculation (implemented by MRCC suite \cite{MRCC}. The Dyall's Hamiltonian\cite{Dyall-ham} is used in spin-free calculation and the spin-dependent calculations are based on Dirac-Coulomb Hamiltonian. In contrast to the spin-free calculation, all $J$-components of the $^3P(3s3p)$ state are obtained as unique eigensolutions in the spin-dependent calculation. The degeneracy (2$L$+1 levels) in the spin-free case, wherein $L$ is the orbital angular momentum quantum number, and the (2$J$+1 levels) degeneracy in the spin-dependent cases are broken to be different $M_L$ and $M_J$ components upon the external perturbation, where $M_{L}$ and $M_{J}$ are the projections of $L$ and $J$ onto the field axis in the $z$ direction. Therefore, the polarizabilities of $^3P$ are obtained for each $M_L$ and $M_J$ component. For spin-dependent case, the dipolarizability for a specific state $|JM_J\rangle$ can be defined as \cite{Mitroy-JPB-2010,Schwerdtfeger-web-2012,Angel-PRPSA-1968,Buchachenkoa-EPJD-2010}
\begin{equation}
\langle JM_J|\alpha_{zz}|JM_J\rangle=\alpha_J(M_J)=\bar{\alpha}^{J}+\alpha_a^J\frac{3M_J^2-J(J+1)}{J(2J-1)},
\end{equation}
where the scalar $\bar{\alpha}^J$ and the tensor polarizabilitieand $\alpha_a^J$ are formulated by
\begin{eqnarray}
\bar{\alpha}^{J}&=&\frac{1}{2J+1}\underset{M_J}{\Sigma}\alpha_{zz}(J,M_J) \\ \nonumber
\alpha_a^{J}&=&\frac{5}{(J+1)(2J+1)(2J+3)}\times \\ \nonumber
& &\underset{M_J}{\Sigma}[3M^2_J-J(J+1)]\alpha_{zz}(J,M_J).
\end{eqnarray}
In the spin-free case, the scalar and tensor polarizabilities $\bar{\alpha}^{L}$ and $\alpha^{L}_{a}$ are also given by Eqs. (2) and (3) with $J$ and $M_J$ replaced by $L$ and $M_L$. The relation between the polarizabilities for different $M_J$ components can be given more explicitly as follows, for $J$ or $L$=1,
\begin{eqnarray}
M_{L,J}=0: &&\alpha(0)=\bar{\alpha}-2\alpha_a \\ \nonumber
M_{L,J}=\pm1:&& \alpha(1)=\bar{\alpha}+\alpha_a,
\end{eqnarray}
and for $J$ or $L$=2,
\begin{eqnarray}
M_{L,J}=0: &&\alpha(0)=\bar{\alpha}-\alpha_a \\ \nonumber
M_{L,J}=\pm1: &&\alpha(1)=\bar{\alpha}-\frac{1}{2}\alpha_a, \\ \nonumber
M_{L,J}=\pm2: &&\alpha(2)=\bar{\alpha}+\alpha_a.
\end{eqnarray}
In the $LS$ coupling approximation one finds \cite{Mitroy-JPB-2010,Schwerdtfeger-web-2012,Angel-PRPSA-1968}
\begin{equation}
\bar{\alpha}^J=\bar{\alpha}^L,
\end{equation}
\begin{eqnarray}
\alpha_a^J=\alpha_a^L(-1)^{S+L+J+2}(2J+1)\left \{ \begin{array}{c c c} S&L&J\\ 2&J&L\end{array}\right\} \\ \nonumber
\times \left ( \begin{array}{c c c} J&2&J\\-J&0&J\end{array}\right )\left ( \begin{array}{c c c} L&2&L\\-L&0&L\end{array}\right ).
\end{eqnarray}
Eqs. (6) and (7) show the relations between the polarizabilites in the $J$ and $L$ representations of an energy level. For the Al$^+\ 3s3p\ ^3P$ state, $L=1$, $S$=1, Eq. (7) can be rewritten as \cite{Mitroy-JPB-2010,Angel-PRPSA-1968}
\begin{equation}
\alpha_a^J(^{3}P_1)=-\alpha_a^J(^{3}P_2)/2=-\alpha_a^L/2.
\end{equation}
The definitions of the scalar and tensor hyperpolarizability are the same as Eqs.(2)-(8) with $\alpha$ replaced by $\gamma$. For the $^1S_0$ and $^3P_0$ states the dipole polarizability and hyperpolarizability have only one component, whereas we remain use $\bar{\alpha}$ and $\bar{\gamma}$ in order to avoid verbose constructions.

The value of the studied properties is convergent with the basis sets of the progressively larger sizes in the CI and CC calculations. We choose the hierarchy of the uncontracted aug-cc-pCVXZ basis set with $X=2, 3, 4$, and 5$\zeta$ \cite{Basis}, where two diffusion functions are added to each shell of the $X=2, 3$, and 4 basis sets, and 2s3p1d1g are added to the $X=5\zeta$ basis set. The details of the basis sets are given in Table I and II. The CI calculations is implemented in the general active space \cite{StefanKnetch} and the details of the CI treatment are illustrated in Table I and II, where `S10' means the single excitation is allowed from 10 core electrons ($1s^22s^22p^6$), `(2in4)SD' means the reference states are generated by 2 valence electrons distributing all possible way in $3s3p$ orbits, allowing the single and double (SD) excitations to the virtual orbits with energy less than a given cutoff (for example, $<100$ a.u.).

The triple excitation into part of virtual orbits (less than 1 a.u.) is considered for the small basis sets $X=2\zeta$ and 3$\zeta$ within spin-dependent CI calculations in order to extract the correction of the triple excitation to polarizibilities. The higher level of electrons correlation is considered within the spin-free CI calculation, which includes the single (S) excitation in the core shell, the single, double, and triple (SDT) excitations from core and valence shells into all virtual orbits. In the spin-dependent CC calculations, the electron correlation of single and double (SD) excitations into virtual orbits with a cutoff 10000 a.u. are considered within the basis set of $X=2, 3$, and 4$\zeta$. The triple excitation is considered for the small basis set $X=2\zeta$ in order to extract the correction of the polarizabilities due to the triple excitation.

In order to present the accurate data of $\bar{\alpha}$ and $\bar{\gamma}$, we adopt the same composite scheme as suggested in Ref.\cite{Kallay-PRA-2011}, which is evaluated by
\begin{equation}
P=P_{\texttt{SD}}+\Delta P_{\texttt{T}}
\end{equation}
where $P$ means the studied properties, $\Delta P_{\texttt{T}}=P_{\texttt{SDT}}-P_{\texttt{SD}}$, $P_{\texttt{SD}}$ and $P_{\texttt{SDT}}$ are the CI or CC calculated values with SD and SDT excitation, respectively. Within the spin-dependent CI calculation the values of $P_{\texttt{SD}}$ and $\Delta P_{\texttt{T}}$ are taken from the $X=5\zeta$ and $X=3\zeta$ basis sets, respectively. Within the spin-dependent CC calculation the values of $P_{\texttt{SD}}$ and $\Delta P_{\texttt{T}}$ are taken from the $X=4\zeta$ and $X=2\zeta$ basis sets, respectively. The error of $P_{\texttt{SD}}$ is computed by $2\times( P_{\texttt{SD}}|_{5\zeta}-P_{\texttt{SD}}|_{4\zeta})$ in the spin-dependent CI calculation and $2\times( P_{\texttt{SD}}|_{4\zeta}-P_{\texttt{SD}}|_{3\zeta})$ in the spin-dependent CC calculation. The error of $\Delta P_{\texttt{T}}$ is roughly estimated with twice itself in both spin-dependent CI and CC calculations. The composite value is determined by Eq. (3) with its error being sum of errors of $P_{SD}$ and $\Delta P_{\texttt{T}}$. The uncertainty of the composite data is assessed by the error divided by the composite data.

\section{results and discussion}
Consider the dipole polarizability first. Table I summarizes the dipole polarizabilities of the Al$^{+}$ ground state $3s^{2}\ ^{1}S_{0}$ and three lower-lying excited state $3s3p\ ^{3}P_{0}$, $^{3}P_{1}$, and $^{3}P_{2}$, as obtained by the different level of electron correlation calculations. The spin-dependent CI and CC calculations give the $J$-resolved polarizability data for each $M_J$ component, then the scalar and tensor polarizabilities, $\bar{\alpha}^J$ and $\alpha_a^J$, are obtained in terms of Eq. (3). The spin-free CI calculation gives the scalar relativistic data of $\bar{\alpha}$ for Al$^+\ 3s^{2}\ ^{1}S_{0}$ and $3s3p\ ^{3}P$ states and $\alpha_a$ for  Al$^+\ 3s3p\ ^{3}P$ state.

\begin{table*}[btp]
\renewcommand{\arraystretch}{1.2}
\setlength{\tabcolsep}{3pt}
\tiny
\caption{Dipole polarizability $\alpha_{zz}$ (a.u.) of Al$^{+}$ }
\begin{tabular}{p{3.9cm} p{0.8cm} c p{0.8cm} c p{0.76cm} p{0.76cm} p{0.76cm} p{0.76cm} c p{0.76cm} p{0.76cm} p{0.76cm} p{0.76cm} p{0.76cm}}\hline\hline
 \multirow{2}{*}{level of excitation}&\multirow{2}{*}{$^{1}S_{0}$} &&\multirow{2}{*}{$^{3}P_{0}$}&&\multicolumn{4}{c}{$^{3}P_{1}$}&\multicolumn{6}{c}{$^{3}P_{2}$}    \\ \cline{6-9}\cline{11-15}
        &            &&              &&$M_J$=0 & $M_J$=1 &$\bar{\alpha}^J$& $\alpha_a^J$ &  & $M_J$=0  & $M_J$=1   & $M_J$=2      &$\bar{\alpha}^J$&$\alpha_a^J$  \\ \hline
 \multicolumn{15}{c}{(a) spin-dependent CI  }  \\
 basis:2$\zeta$(16s,12p,5d)                                 & \multicolumn{14}{c}{ }     \\
 S10(2in4)SD($<$100au)           &23.565 &&24.381 &&24.829&24.217   &24.421  &-0.204 &  &24.126&24.314 &24.878&24.502&0.376  \\
 S10(2in4)SDT($<$1au)SD($<$100au)&23.611 &&24.370 &&24.816&24.207   &24.410  &-0.203 &  &24.117&24.304 &24.865&24.491&0.374  \\
 basis:3$\zeta$(20s,14p,7d,5f)                              & \multicolumn{14}{c}{ }     \\
 S10(2in4)SD($<$100au)           &23.707 &&24.505 &&25.047&24.294   &24.545  &-0.251 &  &24.155&24.390 &25.097&24.626&0.471  \\
 S10(2in4)SD($<$10au)            &23.862 &&24.765 &&25.267&24.576   &24.806  &-0.230 &  &24.459&24.673 &25.317&24.888&0.429  \\
 S10(2in4)SDT($<$1au)SD($<$10au) &23.900 &&24.763 &&25.263&24.575   &24.804  &-0.229 &  &24.459&24.672 &25.313&24.886&0.427  \\
 basis:4$\zeta$(22s17p9d7f5g)                               & \multicolumn{14}{c}{ }     \\
 S10(2in4)SD($<$100au)           &23.784 &&24.231 &&24.891&23.967   &24.275  &-0.30  &  &23.781&24.073 &24.946&24.364&0.582  \\
 basis:5$\zeta$(23s16p10d8f5g3h)                            & \multicolumn{14}{c}{ }     \\
 S10(2in4)SD($<$100au),$P_{CISD}$&23.742 &&24.177 &&24.835&23.906   &24.216  &-0.309 &  &23.700&24.000 &24.883&24.293&0.590  \\
 error in $P_{CISD}$   &$\pm$0.084&&$\pm$0.108&&$\pm$0.112&$\pm$0.122&$\pm$0.118&$\pm$0.018& &$\pm$0.162&$\pm$0.146&$\pm$0.126&$\pm$0.141&$\pm$0.016  \\
 $\Delta P_T$	          &0.038  &&-0.002    &&-0.004    &-0.001      &-0.002  &  0.001   & & 0.000    & -0.001&-0.004    &-0.002    &0.002   \\
 error in $\Delta P_T$ &$\pm$0.076&&$\pm$0.004&&$\pm$0.008&$\pm$0.002&$\pm$0.004&$\pm$0.002& &$\pm$0.000&$\pm$0.002&$\pm$0.008&$\pm$0.004&$\pm$0.004  \\
                                                            &  \multicolumn{14}{c}{  P=$P_{CISD}$+$\Delta P_T$ }              \\
 Composite                       &23.780 &&24.175 &&24.831 &23.905 &24.214  &-0.308  &  &23.700&23.999 &24.879&24.291&0.588  \\
 error                 &$\pm$0.150&&$\pm$0.112&&$\pm$0.120&$\pm$0.124&$\pm$0.122&$\pm$0.020& &$\pm$0.162&$\pm$0.148&$\pm$0.134&$\pm$0.145&$\pm$0.020  \\
 uncertainty($\%$)        &0.63   &&0.46      &&0.48      &0.52      &0.50      &6.49      & &0.68      &0.61     &0.54 &0.60      &3.40    \\\hline
 \multicolumn{15}{c}{(b) spin-dependent CC }  \\
 CCSD ($<$10000a.u.)-2$\zeta$            &24.007  &&24.818& & &24.666& &             &  &24.589&24.765&25.293&24.941&0.352  \\
 CCSDT($<$1000a.u.)- 2$\zeta$	         &23.876  &&24.768& & &24.633& &             &  &24.572&24.732&25.200&24.887&0.313  \\
 CCSD ($<$10000a.u.)-3$\zeta$            &24.164  &&24.977& & &24.770& &             &  &24.637&24.869&25.565&25.101&0.464  \\
 CCSD ($<$10000a.u.)-4$\zeta$,$P_{CCSD}$ &24.238  &&24.656& & &24.632& &             &  &24.182&24.478&25.367&24.774&0.593  \\
 error in $P_{CCSD}$   &$\pm$0.148&&$\pm$0.642&&          &$\pm$0.276&          &    & &$\pm$0.910&$\pm$0.782&$\pm$0.396&$\pm$0.653&$\pm$0.257 \\
 $\Delta P_{5\zeta}\ ^a$                 &-0.042  &&-0.054& & &-0.067& &             &  &-0.081&-0.073&-0.063&-0.071&0.010  \\
 $\Delta P_T$                            &-0.131  &&-0.050& & &-0.033& &             &  &-0.017&-0.033&-0.093&-0.054&-0.039  \\
 error in $\Delta P_T$ &$\pm$0.262&&$\pm$0.100&&          &$\pm$0.060&          &      & &$\pm$0.034&$\pm$0.066&$\pm$0.186&$\pm$0.108&$\pm$0.078  \\
                                                           & \multicolumn{14}{c}{  P=$P_{CCSD}$+$\Delta P_T$}                 \\
 Composite                               &24.065  &&24.552& & &24.532& &             &  &24.084&24.372&25.211&24.650&0.564  \\
 error                 &$\pm$0.410&&$\pm$0.742&&          &$\pm$0.342&          &     & &$\pm$0.944&$\pm$0.848&$\pm$0.582&$\pm$0.760&$\pm$0.178  \\
 uncertainty($\%$)     &1.70      &&3.01      & &         &1.39      & &              & &3.92       &3.47      &2.31       &3.08     &31.56   \\\hline
 \multicolumn{15}{c}{(c) spin-free CI}  \\
                                 &$^1S$                 & & & & & & \multicolumn{5}{c}{$^{3}P$}\\ \cline{5-14}
                                 &                      & & & &  $M_L$=0         & & $M_L$=1        & & &$\bar{\alpha}^L$& &$\alpha_a^L$    \\
S10(2in4)SDT(all orbits)-3$\zeta$& 23.742               & & & & 23.614           & &25.053          & & &24.573          & &0.480            \\
S10(2in4)SDT(all orbits)-4$\zeta$& 23.816$\pm$0.074     & & & &  23.092$\pm$0.52 & &24.880$\pm$0.170& & &24.280$\pm$0.293& &0.600$\pm$0.293   \\ \hline
Ref.[16]                                   &24.140  &&24.622& & &      & &              &  &      &      &      &      &   \\
Ref.[17]                                   &24.137  &&24.614& & &      & &              &  &      &      &      &      &   \\
Ref.[18]                                   &24.048  &&24.543& & &      & &              &  &      &      &      &      &   \\ \hline \hline
\multicolumn{15}{l}{$^a$ $\Delta P_{5\zeta}$ is the correction to the basis set enlarging from $X=4\zeta$ to $5\zeta$ obtained from CISD calculation.}
\end{tabular}
\end{table*}

\begin{table*}[btp]
\renewcommand{\arraystretch}{1.2}
\setlength{\tabcolsep}{3pt}
\tiny
\caption{Hyperpolarizability $\gamma_{zz}$ ($\times10^3$a.u.) of Al$^{+}$ }
\begin{tabular}{p{3.9cm} p{0.8cm} c p{0.8cm} c p{0.76cm} p{0.76cm} p{0.76cm} p{0.76cm} c p{0.76cm} p{0.76cm} p{0.76cm} p{0.76cm} p{0.76cm} }\hline\hline
\multirow{2}{*}{level of excitation}&\multirow{2}{*}{$^{1}S_{0}$}&&\multirow{2}{*}{$^{3}P_{0}$}&&\multicolumn{4}{c}{$^{3}P_{1}$}&\multicolumn{6}{c}{$^{3}P_{2}$}  \\ \cline{6-9}\cline{11-15}
        &            &&              && $M_J$=0& $M_J$=1 &$\bar{\gamma}^J$& $\gamma_a^J$  & & $M_J$=0  & $M_J$=1   & $M_J$=2      &$\bar{\gamma}^J$&$\gamma_a^J$  \\ \hline
\multicolumn{15}{c}{(a) spin-dependent CI  }  \\
basis:2$\zeta$(16s,12p,5d)                                  & \multicolumn{14}{c}{ }     \\

S10(2in4)SD($<$100au)                      &2.411 	  & &6.591 	    &&0.589 	  &9.874 	   &6.779 	    &3.095 	    & &6.544 	   &3.252 	    &0.570 &2.838 	  &-2.473     \\
S10(2in4)SDT($<$1au)SD($<$100au)	       &2.463 	  & &6.591   	&&0.599       &9.932       &6.821       &3.111      & &6.615       &3.284       &0.583 &2.870     &-2.495      \\
basis:3$\zeta$(20s,14p,7d,5f)                               & \multicolumn{14}{c}{ }     \\
S10(2in4)SD($<$100au)	                   &2.651     & &12.186     &&3.871       &16.845      &12.520      &4.325      & &11.303      &6.657       &3.897 &6.482     &-2.905       \\
S10(2in4)SD($<$10au)	                   &2.729     & &11.907 	&&3.990  	  &16.287      &12.188      &4.099      & &12.183      &7.817      &4.032 &7.176      &-3.410       \\
S10(2in4)SDT($<$1au)SD($<$10au)            &2.849     & &11.904     &&4.007       &16.284      &12.192      &4.092      & &12.259      &7.912       &4.029 &7.228     &-3.461       \\
basis:4$\zeta$(22s17p9d7f5g)                                & \multicolumn{14}{c}{ }     \\
S10(2in4)SD($<$100au)                      &2.290  	  & &12.779     &&3.548       &18.074      &13.232      &4.842      & &9.825       &4.551      &3.586 &5.219      &-2.058      \\
basis:5$\zeta$(23s16p10d8f5g3h)                             & \multicolumn{14}{c}{ }     \\
S10(2in4)SD($<$100au),$P_{CISD}$           &2.505     & &13.537     &&3.314       &19.173      &13.887      &5.286      & &8.610       &3.074       &3.536 &4.366     &-1.318       \\
error in $P_{CISD}$                        &$\pm$0.428& &$\pm$1.516 &&$\pm$0.468  &$\pm$2.200  &$\pm$1.622  &$\pm$0.578 & &$\pm$2.430  &$\pm$2.954  &$\pm$0.100  &$\pm$1.706  &$\pm$1.481   \\
$\Delta P_T$	                           &0.120     & &-0.003     &&0.016       &	-0.003     &0.004       &-0.006     & &0.077       &0.095     &-0.003 &0.052       &0.047       \\
error in $\Delta P_T$                      &$\pm$0.24 & &$\pm$0.006 &&$\pm$0.032  &$\pm$0.006  &$\pm$0.0008 &$\pm$0.012 & &$\pm$0.154  &$\pm$0.19  &$\pm$0.006   &$\pm$0.104  &$\pm$0.095  \\
                                                            & \multicolumn{14}{c}{  P=$P_{CISD}$+$\Delta P_T$} \\
Composite                                  &2.625     & &13.534     &&3.330       &19.171      &13.891      &5.280      & &8.687       &3.169       & 3.533      &4.418       &-1.365       \\
error                                      &$\pm$0.648& &$\pm$1.522 &&$\pm$0.50   &$\pm$2.206  &$\pm$1.63   &$\pm$0.59  & &$\pm$2.584  &$\pm$3.144  &$\pm$0.106  &$\pm$1.81   &$\pm$1.576  \\
uncertainty$\%$                            &25.45     & &11.24      &&15.01       &11.5        &11.73       &11.17      & &29.75       &99.21       &3.00 &40.96       &115        \\\hline
\multicolumn{15}{c}{(b) spin-dependent CC }  \\
CCSD($<$10000a.u.)-2$\xi$                  &2.513     & &6.105	    &&            &9.802	   &            &           & &7.141 	   &3.967	    &0.637       &3.270	      &-2.810      \\
CCSDT($<$1000a.u.)-2$\xi$	               &2.523     & &6.473	    &&            &9.245	   &            &           & &7.330 	   &3.636       &0.743       &3.218	      &-2.709     \\
CCSD($<$10000a.u.)-3$\xi$	               &2.881     & &12.146	    &&            &17.129	   &            &           & &11.653 	   &7.057 	    &3.803       &6.675	      &-3.173     \\
CCSD($<$10000a.u.)-4$\xi$,$P_{CCSD}$       &2.538     & &13.728	    &&            &20.682	   &            &           & &9.930 	   &3.683 	    &3.404       &4.821 	  &-1.944     \\
error in $P_{CCSD}$	                       &$\pm$0.686& &$\pm$3.165	&&	          &$\pm$7.106  &	        &           & &$\pm$3.447  &$--^a$ 	&$\pm$0.799	 &$\pm$3.707      &$--$       \\
$\Delta P_T$	                           &0.010     & &0.735	    &&	          &-0.557	   &	        &           & &0.189 	   &-0.330 	    &0.107 	     &-0.052	  &0.101      \\
error in $\Delta P_T$	                   &$\pm$0.020& &$\pm$1.470	&&	          &$\pm$1.147  &	        &           & &$\pm$0.378  &$\pm$0.660 	&$\pm$0.214  &$\pm$0.104  &$\pm$0.202 \\
                                                          & \multicolumn{14}{c}{  P=$P_{CCSD}$+$\Delta P_T$} \\
Composite                  	               &2.548     & &14.463	    &&	          &20.126	   &	        &           & &10.119 	   &3.353	    &3.511       &4.769 	  &-1.843     \\
error               	                   &$\pm$0.786& &$\pm$4.463	&&	          &$\pm$8.253  &	        &           & &$\pm$3.825  &$--$        &$\pm$1.013	 &$\pm$3.811  &$--$       \\
uncertainty($\%$)                          &27.70     & &32.04      &&            &41.01       &            &           & &37.80       &$--$        &28.85       &79.91       &$--$       \\\hline
\multicolumn{15}{c}{(c) spin-free CI}  \\
                                 &$^1S$               & & & & & & \multicolumn{6}{c}{$^{3}P$}\\ \cline{5-14}
                                 &                    & & & & $M_L$=0          & & $M_L$=1        & & &$\bar{\gamma}^L$ & &\multicolumn{2}{l}{$\gamma_a^L$ }    \\
S10(2in4)SDT(all orbits)-3$\zeta$& 2.760              & & & & 19.697           & & 3.905          & & &9.169            & &\multicolumn{2}{l}{-5.264 }          \\
S10(2in4)SDT(all orbits)-4$\zeta$& 2.457$\pm$0.606    & & & & 19.152$\pm$0.544 & & 3.594$\pm$0.311& & &8.780$\pm$0.777  & &\multicolumn{2}{l}{-5.186$\pm$0.293} \\  \hline
Ref.[27]                         &2.368               &\multicolumn{13}{c}{  }                                                           \\ \hline\hline

\multicolumn{15}{l}{$^a$ Here, we fail to estimate the error of $P_{CCSD}$ because of the anomalous value for $^{3}P_{2}$, $M_J$=2 at the basis set of $X$=$3\xi$ }   \\
\end{tabular}
\end{table*}

For Al$^{+}\ 3s^{2}\ ^{1}S_{0}$ and $3s3p\ ^{3}P_{0}$ states, there are already accurate dipole polarizability data available. Mitroy and coworkers have given the first reliable data, $\bar{\alpha}=24.140$ a.u. for $3s^{2}\ ^{1}S_{0}$  and 24.622 a.u. for $3s3p\ ^{3}P_{0}$ \cite{Mitroy-EJPD-2009}. Based on the large basis set up to $X=5\zeta$ and the high-leveled treatment of the electron correlation up to quadruples excitation within the couple-cluster calculations, Kall\"{a}y and coworkers have recommended $\bar{\alpha}=24.137$ a.u. for $3s^{2}\ ^{1}S_{0}$ and $\bar{\alpha}=24.614$ a.u. for $3s3p\ ^{3}P_{0}$ \cite{Kallay-PRA-2011}. Within another calculation that the electron correlation is handled elaborately within the CI plus CC procedures, Safronova and coworkers have recommended $\bar{\alpha}=24.048$ a.u. for $3s^{2}\ ^{1}S_{0}$ and $\bar{\alpha}=24.543$ a.u. for Al$^{+}\ 3s3p\ ^{3}P_{0}$ \cite{Safronova-PRL-2011}. These previously recommended data \cite{Mitroy-EJPD-2009,Kallay-PRA-2011,Safronova-PRL-2011} provide a good benchmark criterion for comparison to prove accuracy of our calculated results.

The quality of the spin-dependent CI results is demonstrated in direct comparison with the spin-dependent CC results. We find an overall trend that the spin-dependent CI values are lower than their corresponding spin-dependent CC values at the same level of the basis set. With the basis set expanding to 5$\zeta$, the spin-dependent CI calculation arrives at the composite value $\bar{\alpha}=23.780$ a.u. for $3s^{2}\ ^{1}S_{0}$ and $\bar{\alpha}=24.175$ a.u. for $3s3p\ ^{3}P_{0}$. These results are within 2\% error, as compared with the previously recommended data \cite{Mitroy-EJPD-2009,Kallay-PRA-2011,Safronova-PRL-2011}.

The electron correlation is more completely considered in the spin-dependent CC calculations, and therefore to prove our accuracy, the most direct comparison is between our CC results and the previously recommended data \cite{Mitroy-EJPD-2009,Kallay-PRA-2011,Safronova-PRL-2011}. Our CC calculation is truncated to the $X=4\zeta$ basis set due to our limited computer power, which may lead to the decrease of accuracy of the CC calculation. However, we find that our CI and CC results change almost the same quantity with increasing basis set, therefore, it is possible to improve our CC results by adding the basis set correction from $X=4\zeta$ to $5\zeta$ obtained from our CI calculation. Finally, our CC results present $\bar{\alpha}=24.065$ a.u. for $3s^{2}\ ^{1}S_{0}$ and $\bar{\alpha}=24.552$ a.u. for $3s3p\ ^{3}P_{0}$. As compared with the previously recommended data \cite{Mitroy-EJPD-2009,Kallay-PRA-2011,Safronova-PRL-2011}, our CC data is most close to Safronova's data \cite{Safronova-PRL-2011} with the agreement up to the first decimal place. This means that our CC data have already arrived sufficient accuracy. However, we have to admit that such truncation of the bases set at $X=4\zeta$ enlarged the uncertainty margins of our calculated results as compared with the previously benchmark calculations \cite{Mitroy-EJPD-2009,Kallay-PRA-2011,Safronova-PRL-2011}.

The overall agreement between our calculated results and the previously recommended data \cite{Mitroy-EJPD-2009,Kallay-PRA-2011,Safronova-PRL-2011} for the Al$^{+}\ 3s^{2}\ ^{1}S_{0}$ and $3s3p\ ^{3}P_{0}$ states gives us confidence for the accuracy of our results for the other two energetically higher lying excited states, $3s3p\ ^{3}P_{1}$ and $^{3}P_{2}$ that have no any recommended data available yet for the best of our knowledge. Here, we expect that the results of $3s3p\ ^{3}P_{1}$ and $^{3}P_{2}$ are of the same precision and reliability because they are obtained together with $3s^{2}\ ^{1}S_{0}$ and $3s3p\ ^{3}P_{0}$ in one calculation with the same energy convergence threshold. The spin-dependent CI calculation arrives at $\bar{\alpha}^{J}=24.214$ a.u. for $3s3p\ ^{3}P_{1}$ and $\bar{\alpha}^{J}=24.291$ a.u. for $3s3p\ ^{3}P_{2}$. The spin-dependent CC calculation yields $\bar{\alpha}^{J}=24.650$ a.u. for $3s3p\ ^{3}P_{2}$ (the $\bar{\alpha}^{J}$ value for $3s3p\ ^{3}P_{1}$ is not obtained because the $M_J=0$ component fails to be found). The tensor polarizability $\alpha_a^J=-0.308$ a.u. (spin-dependent CI value) for $3s3p\ ^{3}P_{1}$, and $\alpha_a^J=0.588$ a.u. (spin-dependent CI value) and 0.564 a.u. (spin-dependent CC value) for $3s3p\ ^{3}P_{2}$. The deviation of $\bar{\alpha}^{J}$ for $3s3p\ ^{3}P_{2}$ state between our CI and CC calculations is less than 1.7\%, within error margins 1.5\% obtained for the ground state $3s^{2}$ $^{1}S_{0}$ and 1.8\% obtained for the lowest-lying excited state $3s3p\ ^{3}P_{0}$. The good agreement in such comparison confirms that our calculations for $3s3p\ ^{3}P_{1}$ and $^{3}P_{2}$ have delivered a good description of the spin-orbit components.

The relativistic effect in the four-components relativistic calculation can be understood as combination of the spin-orbit coupling effect and contraction/decontraction of radial electron density, i.e., the so called scalar relativistic effect. The relativistic effect is discussed through analyzing the $J$-dependence of the scalar and tensor polarizability in this study. In our calculations, the differences $\bar{\alpha}^J(^{3}P_{0})-\bar{\alpha}^J(^{3}P_{1})$ is -0.039 a.u. (spin-dependent CI data), which amounts to only 0.16\% of $\bar{\alpha}^J(^{3}P_{0})$. The difference $\bar{\alpha}^J(^{3}P_{0})-\bar{\alpha}^J(^{3}P_{2})$ is -0.117 a.u. (spin-dependent CI data) and -0.098 a.u. (spin-dependent CC data), which are 0.47\% and 0.40\% of $\bar{\alpha}^J(^{3}P_{0})$, respectively. Such variations of $\bar{\alpha}^J$ for different $J$ components are minor and therefore negligible. The $^{3}P_{0}$ component is of spherically symmetric electron density, and therefore the difference of $\bar{\alpha}^J(^{3}P_{0})$ with respect to the scalar polarizability obtained from the spin-free CI calculation, i.e., $\bar{\alpha}^{L} (^{3}P)- \bar{\alpha}^{J}(^{3}P_{0})$, can be regarded as the impact of the spin-orbit coupling only on the polarizability \cite{Fleig-PRA-2005}. This difference is only 0.105a.u., indicating a weak spin-orbit coupling effect on the dipole polarizability.

Furthermore, the tensor dipole polarizability represents $\alpha_a^J (^{3}P_{1}) \approx -\alpha_a^J (^{3}P_{2})/2 \approx -\alpha_a^L/2$, which is in accordance with Eq. (8). The $J$-resolved $\bar{\alpha}$ and $\alpha_a$ both comply with the basic vector algebra, i.e., Eqs. (6) and (7), implemented under the $LS$ approximation, which reflects a weak impact of the relativity effect on the dipole polarizability of Al$^+\ 3s3p\ ^3P$ state. The $J$ dependence of the dipole polarizability of Al$^+$ is similar to Al atom. In Ref. \cite{Fleig-PRA-2005}, Fleig has found the difference of dipole polarizability between the $J=\frac{1}{2}$ and $J=\frac{3}{2}$ components of Al atom is small, only 0.002 a.u., and therefore Al atom is justified to be essentially nonrelativistic.

Consider the hyperpolarizability. Table II present the results of hyperpolarizability, as computed in the same way with the dipole polarizability. Available hyperpolarizability data are scarcely, because such high-order property is hard to obtained due to more critical computational demand than that for dipole polarizabilities. For the ground state $3s^{2}\ ^{1}S_{0}$, Archibong and Thakkar have obtained $\gamma=2348$ a.u. ( many-body-perturbation theory data). Here, we obtain $\gamma=2625$ a.u. (spin-dependent CI data), 2548 a.u. (spin-dependent CC data), and 2457 a.u. (spin-free CI data), which are 4\%-10\% larger than Archibong and Thakkar's data. This deviation can be attributed to larger basis set and more complete treatment of electron correlation that are used in our calculations.

For the Al$^+\ 3s3p$ excited state, we obtain $\bar{\gamma}^J=13534, 13891$, and 4418 a.u. for $^{3}P_{0}$, $^{3}P_{1}$, and $^{3}P_{2}$ within the spin-dependent CI calculations, and $\bar{\gamma}^J=14463$ and 4769 a.u. for $^{3}P_{0}$ and $^{3}P_{2}$ within the spin-dependent CC calculations (The $\bar{\gamma}^J$ is absent for $^{3}P_{1}$ because its $J=0$ component is not found in our spin-dependent CC calculation). The deviation between the spin-dependent CI and CC results is around 6-8\% (as evaluated by $(\bar{\gamma}^J_{\textrm{CI}}-\bar{\gamma}^J_{\textrm{CC}})/\bar{\gamma}^J_{\textrm{CC}}$), which is within a normal error range consider the hyperpolarizability is very hard to calculate. More comparisons are difficult because there is no data available for Al$^{+}$ $3s3p$ excited state, as the best of our knowledge.

The average of $\bar{\gamma}^J$ of the three $J$ components of the Al$^+\ 3s3p$ excited state, i.e., $[\bar{\gamma}^J(^3P_0)+3\times \bar{\gamma}^J(^3P_1)+5\times \bar{\gamma}^J(^3P_2)]/9$, is closed to the $\bar{\gamma}^L$, which proves some kind of agreement between $\bar{\gamma}^J$ and $\bar{\gamma}^L$. However, $\bar{\gamma^J}$ represents great variations between different $J$ components, which conflicts with Eq. (6). While the difference $\bar{\gamma}^J(^{3}P_{0})-\bar{\gamma}^J(^{3}P_{1})$ is small and therefore negligible, the $\bar{\gamma}^J(^{3}P_{0})-\bar{\gamma}^J(^{3}P_{2})$ is remarkable large, being as much as 67\% of $\bar{\gamma_0}$. The $\bar{\gamma^J}$ result for each component also show more than 50\% deviation from the $\bar{\gamma}^L$ results. With respect to the tensor hyperpolarizability, the ratio $\gamma_a^J(^3P_1)$, $\gamma_a^J(^3P_2)$, and $\gamma_a^L(^3P)$ disagree with the relations given by Eqs. (7) and (8). Considering the convergence of the results with the basis set and the electron correlation level, we think that the numerical error is unlikely to cause such big discrepancy.

The setup of Eqs. (6)-(8) is based on the LS coupling. However, the hyperpolarizability, as high-order responds, is more sensible to the spin-orbit coupling. The mixing of spin and spatial degrees of freedom leads to deviations from the purely spatial anisotropies. This may cause deviation in the hyperpolarizabilities of light atoms from Eqs. (6)-(8) and dipole polarizabilities of the heavy atoms. The latter has already been founded for In and Tl atoms \cite{Fleig-PRA-2005}. The discrepancy shown in our data for Al$^+$ $3s3p$ excited state, as compared with Eqs. (6)-(8), indicates that the hyperpolarizability is still of open questions especially for the excited state. Currently, there are very few hyperpolarizability data for the excited state, even simple atom, therefore more calculations of high accuracy are needed on the future.

One important application of the scalar polarizabilities is to determinate the BBR shift for a transition due to the finite background thermal radiation. For Al$^{+}$, the BBR shift of the transition $^{1}S_0$ and $^3P_0$ is of especially important meanings for assessing the systematic error of the clock-frequency measurement. The derivation of the theoretical BBR shift has been presented by Porsev and coworkers \cite{Porsev-PRA-2006} and Arora and coworkers \cite{Arora-PRA-2012}, which has shown that the dominant term of BBR is determined by the difference in the dipole polarizability as follows,
\begin{equation}
\delta E^{E_1}=-\frac{1}{2}\frac{4\pi^3\alpha^3}{15}(k_BT)^4\Delta\bar{\alpha}(1+\eta).
\end{equation}
where $E_1$ means the first-order channel in electric field, the fine-structure constant $\alpha=1/137.035 999 074(44)$, and $(\frac{k_BT}{E_h})\approx10^{-9}$ at room temperature the temperature, $\Delta \bar{\alpha}$ means difference in $\bar{\alpha}$. The parameter $\eta$ has been calculated by Mitroy, et al. \cite{Mitroy-EJPD-2009} and Safronova, et al. \cite{Safronova-PRL-2011}, which gives $\eta$=0.00022$\sim$0.00024 for Al$^{+}$. In this paper, we do not calculate this value. Consider that $\eta$ is very small, we therefore neglect this value in the our following estimation of BBR shifts. The above equation can be rewritten as
\begin{equation}
\delta E^{E_1}=-\frac{1}{2}\Delta\bar{\alpha}\langle E^2_{E_1}\rangle,
\end{equation}
where the electric field $\langle E^2_{E_1}\rangle$ is equivalent to $F_z^2$ shown in Eq. (1). By associate the high-order term in Eq. (1), we suppose that the contribute of the hyperpolarizability to the BBR shift can written in an approximated way as
\begin{equation}
\delta E^{E_1}=-\frac{1}{24}\Delta\bar{\gamma}\langle E^2_{E_1}\rangle ^{2},
\end{equation}
where $\Delta \bar{\gamma}$ is differential hyperpolarizability between two states. Based on the scalar polarizability data shown in Table I and II, $\Delta \bar{\alpha}$ and $\Delta \bar{\gamma}$ between Al$^{+}\ 3s^2\ ^{1}S_{0}$ and $3s3p\ ^{3}P_J$ with $J=0,1,2$ and their corresponding BBR shifts can be computed in terms of Eq. (4)-(6), as given in Table III. Such results show that the BBR shifts caused by the hyperpolarizability is of a factor of $10^{-17}$, which is far less than the case of dipole polarizability, and therefore which will constitutes no impediment to the accuracy of the Al$^+$ optical clock at $10^{-18}$ and even higher precisions.

\begin{table*}[btp]
\renewcommand{\arraystretch}{1.2}
\setlength{\tabcolsep}{3pt}
\scriptsize
\caption{Differential dipole polarizability $\Delta \bar{\alpha}$, differential hyperpolarizabity $\Delta \bar{\gamma}$, and BBR shifts $\Delta v^a$}
\begin{tabular}{l l l l l l l l l  l l }\hline\hline
Transition& &$\Delta \bar{\alpha}$(a.u.)& & $\Delta v_{\Delta \bar{\alpha}}$($\times 10^{-3}$Hz)& &$\Delta \bar{\gamma}$ ($\times 10^4$a.u.)& &$\Delta v_{\Delta \bar{\gamma}}$ ($\times 10^{-17}$Hz)$^b$   & & Source \\\hline
($^{1}S_{0}$--$^{3}P_{0}$)& &0.39$\pm$0.038  & &-3.334$\pm$0.324  & &1.091$\pm$0.956    & &-2.02 $\pm$1.771      & & KRCI  \\
                          & &0.487$\pm$0.332 & &-4.163$\pm$2.838  & &1.192$\pm$0.324    & &-2.209$\pm$0.600      & & MRCC  \\
                          & &0.48$\pm$0.125  & & -4.2$\pm$3.2     & &                   & &                     & & Ref.[16]  \\
                          & &0.477$\pm$0.078 & & -4.1$\pm$0.7     & &                   & &                     & & Ref.[17]  \\
                          & &0.495           & & -4.26$\pm$0.43   & &                   & &                     & & Ref.[18]  \\\hline
($^{1}S_{0}$--$^{3}P_{1}$)& &0.434$\pm$0.028 & &-3.71$\pm$0.239   & &1.127$\pm$0.113    & &-2.924$\pm$0.209      & & KRCI  \\\hline
($^{1}S_{0}$--$^{3}P_{2}$)& &0.508$\pm$0.006 & &-4.342$\pm$0.051  & &0.179$\pm$0.127    & &-0.332$\pm$0.235      & & KRCI  \\
                          & &0.585$\pm$0.356 & &-5.001$\pm$3.043  & &0.221$\pm$0.306    & &-0.410$\pm$0.567      & & MRCC  \\\hline\hline
\multicolumn{11}{l}{$^a$ BBR shift is evaluated at temperature $T$=300K. $^b$ $\Delta E\sim-\frac{1}{24}\langle E^{2}_{E1}(\omega)\rangle^2 \Delta\gamma$ is assumed.}                                    \\
\end{tabular}
\end{table*}

\section{summary}
The accurate dipole polarizability and hyperpolarizablilty have been achieved for Al$^{+}\ 3s^2\ ^{1}S_{0}$ and $3s3p\ ^{3}P_J$ with $J=0,1,2$ using relativistic Dirca-Coulomb Hamiltonian within CI and CC theories and a finite-field approach. Our calculations have obtained the accurate dipole polarizabilities, more importantly we present the $J$-dependence and anisotropy of the dipole polarizability and hyperpolarizability. Because of the large computation demanding in finite-field study of the polarizabilities, we do not pursue the highest accuracy, for example, within the spin-dependent CI calculation the single and double electron correlations are limited to virtual orbit less than 100 a.u. and the single, double, and triple electron correlations are limited to virtual orbits less than 1 a.u. Within the spin-dependent CC calculation we truncated the increasing bases set up to $X=4\zeta$. Such truncations of the electron correlation and the basis set cause the increased uncertainty, as compared with the previously benchmark calculations \cite{Kallay-PRA-2011,Safronova-PRL-2011}.

There are more sources of error, such as the correction of quadruples excitation $P_Q$ and the Briet interaction and QEC correction $P_{BQ}$ \cite{Kallay-PRA-2011}. In the previously benchmark calculation the changes in dipole polarizability due to $P_Q$ and $P_{BQ}$ are found to be less than 0.1\% and 1\%, respectively. Therefore, the error caused by missing of $P_Q$ and $P_{BQ}$ should not exceed a factor $1\sim2$\% for our calculated results. The changes of the hyperpolarizability due to $P_Q$ and $P_{BQ}$ should be a small correction in the similar trend shown in the dipole polarizability.

Though such imperfect in our calculation, our results have shown the excellent agreement with previously recommended data the dipole polarizability of $3s^{2}\ ^{1}S_{0}$ and $3s3p\ ^{3}P_{0}$ and the hyperpolarizability of $3s^{2}\ ^{1}S_{0}$ as well as the excellent agreement between the spin-dependent CI and CC calculations. It is indicated that the spin-orbit coupling has negligible contribution for the dipolarizability of Al$^+$ but become a significant for the hyperpolarizability. Therefore, the fully relativistic calculation is strongly demanded for the high-order polarizabilities.

Finally, we evaluated the BBR shift due to dipole polarizability and hyperpolarizability for Al$^{+}\ 3s^{2}\ ^{1}S_{0}$ to $3s3p\ ^{3}P_{J}$ with $J=0,1,2$. Specially, the BBR shifts caused by the hyperpolarizability is at the magnitude of $10^{-17}$Hz, which is far lower than the precision level of the current Al$^+$ optical clock and therefore can be safely neglected in uncertainty budget.

\section{Acknowledgements}
The authors thank Dirac experts for helpful discussion through the Dirac mailing list. The authors are grateful to Prof. Zong-Chao Yan and Prof. Jim Mitroy for helpful suggestions.
This work is supported by 2012CB821305, 2010CB922904, NSFC 61275129, NFSC 21033001, NFSC 21203147, and CAS KJZD-EW-W02.

\newpage

\begin{thebibliography}{}

\bibitem{Itano-JRNIST-2000}
W. M. Itano, J. Res. Natl. Inst. Stand. Technol. 105, 829 (2000).

\bibitem{Rosenband-arXiv-2000}
T. Rosenband, W. M. Itano, P. O. Schmidt, D. B. Hume, arXiv:physics/0611125.

\bibitem{Mitroy-JPB-2010}
J. Mitroy, M. S. Safronova and C. W. Clark, J. Phys. B: At. Mol. Opt. Phys {\bf 43} 1 (2010).

\bibitem{Safronova-IEEE-2012}
M. S. Safronova, M. G. Kozlov, and C. W. Clark, IEEE Trans. Ultrasonics, Ferroelectrics, and Frequency Control {\bf 59} 439 (2012).

\bibitem{Porsev-PRA-2006}
S. G. Porsev and A. Derevianko, Phys. Rev. A {\bf 74}, 020502(R) (2006).

\bibitem{Arora-PRA-2012}
B. Arora, D. K. Nandy, and B. K. Sahoo, Phys. Rev. A {\bf 85}, 012506 (2012).

\bibitem{Lim-PRA-2004}
I. S. Lim and P. Schwerdtfeger, Phys. Rev. A {\bf 70} 062501(R) (2004).

\bibitem{Fleig-PRA-2005}
T. Fleig, Phys. Rev. A {\bf 72} 052506 (2005).

\bibitem{Lupinetti-JCP-2005}
C. Lupinetti, and A. J. Thakkar, J. Chem. Phys. {\bf 122} 044301 (2005).

\bibitem{Miller-CRC-2007}
T. M. Miller, Atomic and Molecular Polarizabilities, CRC Press, Boca Raton, Florida, 2007, vol. 88, chap. 10, pp. 101-192

\bibitem{Thiefelder-PRA-2008}
C. Thierfelder, B. Assadollahzadeh, P. Schwerdtfeger, S. Sch\"{a}fer, R. Sch\"{a}fer, Phys. Rev. A {\bf 78} 052506 (2008).

\bibitem{Schwerdtfeger-web-2012}
The CTCP table of experimental and calculated static dipole polarizabilities for the electronic ground states of the neutral elements, http://ctcp.massey.ac.nz/dipole-polarizabilities.

\bibitem{Sahoo-PRA-2012}
B. K. Sahoo and B. P. Das, Phys. Rev. A {\bf 86} 022506 (2012).

\bibitem{Porsev-PRA-2012}
S. G. Porsev, M. S. Safronova, and M. G. Kozlov, Phys. Rev. A {\bf 85} 062517 (2012).

\bibitem{Chou-PRL-2010}
C. W. Chou, D. B. Hume, J. C. J. Koelemeij, D. J. Wineland, and T. Rosenband, Phys. Rev. Lett. {\bf 104} 070802 (2010).

\bibitem{Mitroy-EJPD-2009}
J. Mitroy, J. Y. Zhang, M. W. J. Bromley, and K. G. Rollin, Eur. Phys. J. D {\bf 53}, 15(2009).

\bibitem{Kallay-PRA-2011}
M. Kall\"{a}y, H. S. Nataraj, B. K. Sahoo, B. P. Das, and L. Visscher, Phys. Rev. A {\bf 83} 030503(R) (2011).

\bibitem{Safronova-PRL-2011}
M. S. Safronova, M. G. Kozlov, and C. W. Clark, Phys. Rev. Lett. {\bf 107} 143006 (2011).


\bibitem{Angel-PRPSA-1968}
J. R. P. Angel and P. G. H. Sandars, Proc. Roy. Phys. Soc. A 305, 125 (1968).

\bibitem{McLean-JCP-1967}
A. D. McLean and M. Yoshimine, J. Chem. Phys. {\bf 47} 1927 (1967).

\bibitem{Thakkar-JCP-1988}
G. Maroulis and A. J. Thakkar, J. Chem. Phys. {\bf 88} 7623 (1988).

\bibitem{Safronova-PRA-2012}
Z. Zuhrianda, M. S. Safronova, and M. G. Kozlov, Phys. Rev. A {\bf 85} 022513 (2012).

\bibitem{Safronova-PRA-2013-Sr}
M. S. Safronova, S. G. Porsev, U. I. Safronova, M. G. Kozlov, and Charles W. Clark, Phys. Rev. A {\bf 87} 012509 (2013).

\bibitem{Chen-PRA-2012}
Y. Cheng and J. Mitroy, Phys. Rev. A {\bf 86} 052505 (2012).

\bibitem{Chen-JPB-2013}
Y. Cheng and J. Mitroy, J. Phys. B: At. Mol. Opt. Phys. {\bf 46} 185004 (2013).

\bibitem{Cheng-PRA-2013}
Y. Cheng, J. Jiang, and J. Mitroy, Phys. Rev. A {\bf 88} 022511 (2013).

\bibitem{Archibong-PRA-1991}
E. F. Archibong and A. J. Thakkar, Phys. Rev. A {\bf 44} 5478 (1991).

\bibitem{Dirac}
DIRAC, a relativistic ab initio electronic structure program, Release DIRAC11 (2011), written by R. Bast, H. J. Aa. Jensen, T. Saue, and L. Visscher

\bibitem{MRCC}
MRCC, a string-based quantum chemical program suite written by M. Kallay. See also M. Kallay, P. R. Surjan, J. Chem. Phys. 115, 2945 (2001) as well as: www.mrcc.hu.

\bibitem{Dyall-ham}
K. G. Dyall, J. Chem. Phys. 100, 2118 (1994).

\bibitem{Buchachenkoa-EPJD-2010}
A. A. Buchachenkoa, Phys. J. D 61, 291, (2011).

\bibitem{Basis}
D. E. Woon and T.H. Dunning, Jr.  J. Chem. Phys. 98, 1358 (1993).

\bibitem{StefanKnetch}
Stefan Knecht, Parallel Relativistic Multiconfiguration Methods: New Powerful Tools for Heavy-Element Electronic-Structure Studies, HHU D¨¹sseldorf, D, (2009).










\end{thebibliography}
\end{document}